\documentclass[pra, aps,superscriptaddress, twocolumn]{revtex4-1}

\usepackage{subfigure}

\usepackage{dcolumn}
\usepackage{amsmath,amssymb}
\usepackage{bm}
\usepackage{color}
\usepackage{overpic}
\usepackage{latexsym}
\usepackage{epstopdf}
\usepackage{color}
\usepackage[english]{babel}
\usepackage{graphicx}
\usepackage{psfrag} 
\usepackage{epsf} 
\usepackage{subfigure} 
\usepackage{amsfonts}
\usepackage{bm}
\usepackage{natbib}
\usepackage{epstopdf}\usepackage{appendix}

\definecolor{mygrey}{gray}{0.35}
\definecolor{myblue}{rgb}{0.2,0.2,0.8}
\definecolor{myzard}{cmyk}{0,0,0.05,0}
\definecolor{mywhite}{rgb}{1,1,1}
\definecolor{myred}{rgb}{1,0.,0.3}

\usepackage[colorlinks=true,citecolor=myblue,linkcolor=myred]{hyperref}
\def\beq{\begin{equation}}
\def\eeq{\end{equation}}
\def\ba{\begin{align}}
\def\enda{\end{align}}
\def\bi{\begin{itemize}}
\def\ei{\end{itemize}}

\def\beq{\begin{equation}}
\def\eeq{\end{equation}}

 \newcommand{\ket}[1]{|#1\rangle}
 \newcommand{\bra}[1]{\langle #1|}

 \newcommand{\ketbra}[2]{\ket{#1}\bra{#2}}
 \newcommand{\proj}[1]{\ketbra{#1}{#1}}

\graphicspath{{./figs/}}

\begin{document}


\title[]{Noise-induced transport in the motion of trapped ions}

\author{Cecilia Cormick}
\affiliation{Instituto de F\'isica Enrique Gaviola, CONICET and Universidad Nacional de C\'ordoba, Ciudad Universitaria,
X5016LAE, C\'ordoba, Argentina}

\author{Christian T. Schmiegelow}
\affiliation{Departamento de F\'isica, FCEyN, UBA and IFIBA, CONICET, Pabell\'on 1, Ciudad Universitaria, 1428 Buenos Aires, Argentina}

\date{\today}

\begin{abstract}
The interplay of noise and quantum coherence in transport gives rise to rich dynamics relevant for a variety of systems. In this work, we put forward a proposal for an experiment testing noise-induced transport in the vibrational modes of a chain of trapped ions. We focus on the case of transverse modes, considering multiple-isotope chains and an ``angle trap'', where the transverse trapping varies along the chain. This variation induces localization of the motional modes and therefore suppresses transport. By suitably choosing the action of laser fields that couple to the internal and external degrees of freedom of the ions, we show how to implement effective local dephasing on the modes, broadening the vibrational resonances. This leads to an overlap of the local mode frequencies, giving rise to a pronounced increase in the transport of excitations along the chain. We propose an implementation and measurement scheme which require neither ground-state cooling nor low heating rates, and we illustrate our results with a simulation of the dynamics for a chain of three ions.
\end{abstract}

\maketitle

\section{Introduction}

The observation of quantum coherences in the process of excitonic transport in photosynthetic complexes \cite{Engel_Nature_2007, Panitchayangkoon_PNAS_2010} has fostered the analysis of the roles of coherence and noise in transport processes in general. While coherent dynamics may lead to an improvement of transport times due to superradiance and similar phenomena \cite{Struempfer_JPCL_2012}, it has also been shown that in simplified models of photosynthetic complexes a certain noise level can be beneficial to suppress destructive interferences or compensate energy differences that would lead to localization of excitations \cite{Chin_PhilTrans_2012, Chin_NatPhys_2013}. The phenomenon has been analyzed in various systems, showing that transport can be favoured by noise in ultracold atoms \cite{Mendoza-Arenas_2013}, chains of quantum dots \cite{Debora_2014}, or condensed matter setups \cite{Ghosh_2005}. 

Ion chains constitute an appealing platform for the experimental exploration of complex dynamics due to their high degree of controlability. For example, chains of few ions in linear Paul traps have been used to simulate Ising spin chains~\cite{Kim_2010,Senko_2015}, open-system dynamical maps~\cite{Schindler_2013}, and quasiparticle dynamics in many-body systems~\cite{Jurcevic_2014}. In Penning traps, controlled spin dynamics have recently been realized with hundreds of trapped ions~\cite{Bohnet_2016}. Other studies have focused on thermodynamic behaviour, demonstrating thermalization of a spin coupled to a bosonic environment~\cite{Clos_2015} and energy transfer along a chain of up to 37 ions~\cite{Ramm_2014}. 

The richness of the dynamics of trapped-ion systems can be further expanded through the addition of optical potentials to controllably modify the spatial structure of the crystal and its motional modes \cite{Horak_2012}. Indeed, dipole potentials have been proposed and used for the investigation of friction models~\cite{Garcia-Mata_2007, Benassi_2011, Bylinskii_2015, Pruttivarasin_2011}. Similar experimental techniques have been suggested for 
the generation of coherent superpositions of motional states of small crystals \cite{Baltrusch_2011}, and for simultaneous cavity cooling of all axial modes of an ion chain \cite{Fogarty_2016}. The proposed extension of nonlinear spectroscopy to ion traps can enable the measurement of small couplings and nonlinear dynamics of motional excitations~\cite{Lemmer_2015, Schlawin_2014, Gessner_2014}. An essential requirement for these schemes, the realization of phase-stable and controllable standing waves, has been recently demonstrated~\cite{Schmiegelow_2016}.

Previous studies of transport of vibrational excitations in ion traps have analyzed the role of the spatial structure, the impact of disorder~\cite{Freitas_2015} and the onset of Fourier's law as a result of disorder or dephasing \cite{Bermudez_PRL_2013}. Furthermore,  a procedure to measure heat transport was presented in \cite{Bermudez_PRL_2013}. However, the proposal in \cite{Bermudez_PRL_2013} does not offer a simple and scalable way to introduce local and time-dependent energy fluctuations. We now take a crucial step further by introducing a method for the experimental implementation of local dephasing noise on the vibrational excitations by means of fluctuating optical potentials. 

\begin{figure}[t]
\includegraphics[width=0.85\columnwidth]{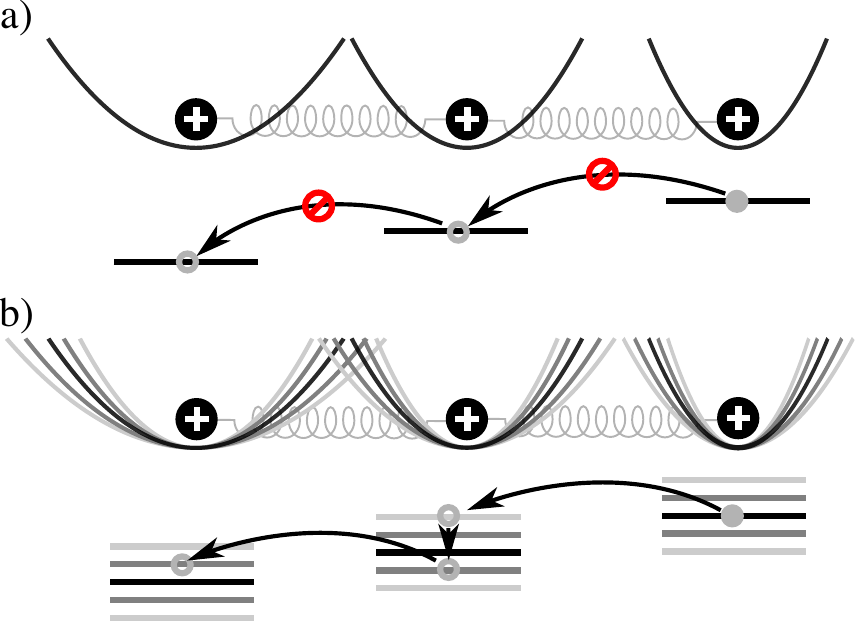}
\caption{
\label{fig:sketch}
Sketch of ions in an inhomogeneous trap, together with the energy levels of their vibrational states. a) If the differences in local frequency are large compared with the couplings, there is no energy transfer along the chain. b) The addition of local frequency fluctuations allows excitations to resonantly hop from site to site.
}
\end{figure}

As an application of our idea, we show how it can be used to observe noise-induced transport in a system which would otherwise exhibit localized excitations (see Fig. \ref{fig:sketch}). An example of such a scenario is provided by a so-called ``angle trap'', i.e. a radiofrequency trap with a transverse trapping frequency which depends on the axial position of the ion. Such traps have already been built and used for the realization of small thermal machines \cite{heat-engine-1, heat-engine-2, heat-engine-3}. Alternatively, localization of transverse modes can be achieved trapping different isotopes of a given ion species, since the transverse trapping frequency depends on the mass \cite{Leibfried_2003}.

The article is organized as follows: In Sec. \ref{sec:angle trap} we describe vibrational dynamics in the weak coupling regime, in which the variation of transverse trapping leads to localization of the vibrational excitations, and provide realistic parameters for angle-trap and multi-isotope ion chains. Section \ref{sec:ideas} discusses different possible strategies to observe noise-induced transport, including an analysis of drawbacks of previous proposals. In Sec. \ref{sec:noise} we explain our procedure to introduce local frequency fluctuations by means of optical fields, while in Sec. \ref{sec:measurement} we discuss how to detect the transport of vibrations. Section \ref{sec:results}~shows the expected results of an implementation with a small chain of three ions. A final discussion is included in Sec. \ref{sec:conclusions}.

\section{Transverse vibrational dynamics and mode localization}
\label{sec:angle trap}

In this Section we describe the transverse vibrational dynamics of a chain of singly charged ions in cases where the normal modes can be localized. We treat the potential in harmonic approximation about the equilibrium positions of the ions, and assume that the frequencies in the three directions are different enough that we can analyze the motion in only one transverse direction, which we call $x$. We focus on the regime where the transverse frequencies are much larger than the Coulomb couplings between ions, so that the interaction terms can be described as hopping of one vibrational excitation from one ion to another. 

We consider two simple experimental setups where one can observe localization of the transverse modes in an ion trap. The first one is an angle trap, that is, a trap with electrodes forming an angle, such that the transverse trapping frequency depends on the axial coordinate $z$ of the ion~\cite{Tolazzi-thesis}. Alternatively, one can use a chain with different isotopes of the same ion species, resulting in small differences in the transverse trapping frequency experienced by each isotope \cite{Leibfried_2003}. In both cases, choosing suitable trapping parameters one can make the differences in transverse frequencies much larger that the coupling constants, so that the transverse vibrational modes of the chain are localized. 

\subsection{Transverse motion in an angle trap}

We consider first the case of ions of equal mass in a trap in which the transverse frequency depends on the axial position of the ion. 
Such a trap was proposed as a single-ion heat engine~\cite{heat-engine-1, heat-engine-2}, and its experimental demonstration has been recently published~\cite{heat-engine-3}; details about the trap design and features can be found in~\cite{Tolazzi-thesis}. 

Formally, the potential for the transverse degrees of freedom can be written as $
V = V_0 + V_{\rm int}$, where $V_0$ contains the terms of the potential that depend separately on the different ions, 
\beq
V_0 = \sum_{j=1}^N \frac{m \omega_j^2}{2} x_j^2; 
\eeq
with $x_j$ the transverse coordinate of the $j$-th ion and $\omega_j$ the local transverse frequency. This local frequency includes a renormalization due to the Coulomb interaction with the other ions in the form:
\beq
\omega_j^2 = \omega_{x,j}^2 - \omega_z^2 \sum_{k\neq j} \left(\frac{l}{d_{jk}}\right)^3.
\label{eq:renormalization}
\eeq
Here, $\omega_{x,j}$ is the transverse trap frequency for ion $j$ and $l$ is a distance unit given by:
\beq
l = \left(\frac{e^2}{4 \pi \epsilon_0 m \omega_z^2} \right)^{1/3}.
\label{eq:l}
\eeq
The term $V_{\rm int}$ contains the Coulomb-interaction terms between ions, 
\beq
V_{\rm int} = m \omega_z^2 \sum_{j < k} \left(\frac{l}{d_{jk}}\right)^3 x_j x_k.
\label{eq:Coulomb}
\eeq

Defining the usual creation and annihilation operators $a_j, a_j^\dagger$ associated with the position and momentum operators $x_j, p_j$, and considering that the coupling strengths are much weaker than the transverse trapping frequencies, one gets an effective Hamiltonian for the modes in the form: $H = H_0 + H_c$. Here, $H_0$ contains local terms,
\beq
H_0 =  \hbar \sum_{j=1}^N \omega_j a^\dagger_j a_j ,
\eeq
and $H_c$ couples sites describing tunneling of vibrational excitations from one ion to the other due to their Coulomb interaction:
\beq
H_c = \hbar \sum_{j < k} c_{jk} (a^\dagger_j a_k + a^\dagger_k a_j )\,,
\label{eq:Hc}
\eeq
where the coupling constants take the form:
\beq
c_{jk} = \frac{1}{2} \frac{\omega_z^2}{\sqrt{\omega_j\omega_k}} \left( \frac{l}{d_{jk}} \right)^3.
\label{eq:couplings}
\eeq
The present form of the Hamiltonian is valid as long as $|c_{j,k}| \ll \omega_x$ with $\omega_x$ a typical transverse frequency. The excitations that diagonalize this Hamiltonian are localized in the regime when $|c_{j,k}| \ll |\omega_j-\omega_k|$ for all pairs of ions. This means that the degree of localization of the normal modes depends on how the local frequency differences compare with $\omega_z^2/\omega_x$.

We now discuss some parameters which are relevant for our proposal, based on the trap described in \cite{Tolazzi-thesis}. Axial trapping frequencies for $^{40}$Ca$^+$ in this trap are typically in the order of 10-500 kHz, while transverse frequencies can be chosen in the range from 200 to 600 kHz. The key trap feature is the dependence of the transverse trapping on the axial coordinate. Under the previous conditions, this trap exhibits a variation of about 10\% in the transverse frequency for a axial displacement of about 250 $\mu$m. For a crystal with three $^{40}$Ca$^+$ ions and an axial trapping frequency of 40 kHz, leading to a distance between ions of about 40 $\mu$m, and with transverse frequencies around 440 kHz, differences of approximately 5 kHz are measured between neighbouring ions. For these parameters, localization of the transverse modes has been experimentally demonstrated.  

When modes are localized, transport of vibrational excitations along the chain is suppressed: it is possible to excite the motion of one ion at one end of the chain without significantly affecting the vibrational state of an ion at the other end of the chain. This is in strong contrast with the situation in a homogeneous trap, where motional modes are delocalized and have support over the chain as a whole, thus leading to transport of excitations. Increasing the axial trapping frequency in the angle trap, the ions can be brought closer to each other and the difference in the transverse trapping experienced by the different ions is reduced while coupling strengths are increased. This in turn suppresses the localization of the modes, which for small distances become wave-like, approximating the delocalized modes of a homogeneous trap. 

\subsection{Transverse modes in a multi-isotope chain}

In a standard linear trap, the desired differences in local transverse frequency can be achieved using different isotopes of the same ion species. This gives a limited number of possibilities, but we note that Yb$^+$ has seven observationally stable isotopes (five of them with no nuclear spin), and is a species of frequent use in ion-trap setups. For a transverse frequency of about 500 kHz, the frequency difference between consecutive stable isotopes would be of about 3-6 kHz, which is of the same order as the frequency variations reported in the previous subsection for representative values in the angle trap.

A detailed analysis of the modes for chains of ions with different masses can be found for instance in \cite{Morigi_Walther_2001}. To use a notation similar to the one in the previous subsection, we now choose a reference isotope of mass $m_0$ to define reference trap frequencies $\omega_{z,0}$ and $\omega_{x,0}$. The definition of the length scale in Eq.~(\ref{eq:l}) is then modified accordingly, replacing $m\omega_z^2$ by the reference values. The equilibrium positions of the ions are not affected by their masses, and so the Coulomb interaction takes the same form as in Eq.~(\ref{eq:Coulomb}), with the same replacement for $m\omega_z^2$. The pseudopotential experienced by each ion depends on its mass, resulting in a transverse trap frequency for each ion of the form \cite{Leibfried_2003}:
\beq
\omega_{x,j} = \frac{m_0}{m_j} \omega_{x,0}.
\label{eq:transverse frequency}
\eeq
The equations that describe the local frequency shift due to the Coulomb interaction, and the interaction terms between ions, Eqs.~(\ref{eq:renormalization}) and (\ref{eq:Coulomb}), must also be modified replacing $\omega_z$ and $m$ by $\omega_{z,0}$ and $m_0$. In this way all equations from the previous subsection are generalized to the multi-isotope case.

One then defines local creation and annihilation operators for the transverse motion taking into account the different masses and (renormalized) local frequencies. The final Hamiltonian takes the same form as in the previous subsection but with the coupling coefficients $c_{jk}$ in Eq.~(\ref{eq:Hc}) now given by:
\beq
c_{jk} = \frac{1}{2} \frac{\omega_z^2}{\sqrt{\omega_j\omega_k}} \frac{m_0}{\sqrt{m_j m_k}} \left( \frac{l}{d_{jk}} \right)^3.
\eeq
We note that the inverse dependence on the mass of the local trap frequency, Eq. (\ref{eq:transverse frequency}), does not lead to a simplification of this formula unless the renormalization shift is negligible. As before, the form of the coupling Hamiltonian (\ref{eq:Hc}) is obtained neglecting counter-rotating terms which do not conserve the total number of local excitations.

\section{Noise-induced transport}
\label{sec:ideas}

The main idea of the present proposal is to study the transition between a regime of localized excitations to one where transport is possible due to the action of external noise. Indeed, local dephasing of the transverse modes can broaden the resonances and bridge the gap between neighbouring ions, thus allowing the excitations to tunnel across sites (each site corresponding to a particular ion).
In this way, transport properties of the chain are altered by means of additional dephasing noise, which can be introduced coupling the transverse motion with internal degrees of freedom. This idea was already contained in \cite{Bermudez_PRL_2013}, but the two methods considered there for the realization of the dephasing noise are not suitable for the kind of dynamics we want to analyze. 

The first procedure proposed in \cite{Bermudez_PRL_2013} was the addition of electric noise to the electrodes, leading to fluctuations in the transverse trapping. In order to have some impact on the transport trapping, these fluctuations should not be global. This requires working near the surface of segmented traps which can produce potentials that vary significantly over the scale of the interparticle distance. However, because of the ion proximity to the surface, this procedure comes at the expense of additional motional heating~\cite{Brownnutt_2015}.

The second strategy proposed in \cite{Bermudez_PRL_2013} was the introduction of disorder in the motional frequencies using a coupling to the internal states in the form:
\beq
H_{\rm int} = \sum_j \Delta \omega a^\dagger_j a_j \sigma^z_j
\label{eq:frequency shift}
\eeq
and taking an initial state where each ion is put in a superposition of the two internal states. The implementation of this Hamiltonian requires two non-copropagating laser beams inducing two-photon processes involving absorption and emission into different beams. An initial superposition state like the one considered will give rise to a distribution of frequencies in which each local frequency is modified in the form $\omega_j \to \omega_j \pm \Delta \omega$ where the sign is random and all realizations are averaged. This can indeed modify the transport properties in the chain, as shown in \cite{Bermudez_PRL_2013}. Nevertheless, the disorder introduced in this way is static, and in general will not be sufficient to prevent localization of excitations in the scenarios we consider.

We are interested in modifying these ideas to obtain a dynamical broadening of the resonances that can induce transport of otherwise localized vibrational excitations. We now discuss some alternatives:

1) One can implement a coupling of the form of the above Hamiltonian (\ref{eq:frequency shift}), but leaving all the spins in the ground state, and instead making the quantity $\Delta \omega$ site dependent, i.e. $\Delta \omega \to \Delta \omega_j$. This might be done if the beams are not plane waves, but have a spatial intensity gradient along the chain axis (taking advantage of the fact that typical ion distances are much longer than a wavelength). For the disorder to be dynamical, one needs to introduce during the experiment pseudo-random axial shifts of this wavefront, so that the frequency differences between neighbouring ions vary during the experiment. This requires having a fluctuating optical field, negligible forces on the axial direction, and optical frequency shifts comparable with the frequency gap between neighbours. 

2) One can implement exactly the Hamiltonian proposed in \cite{Bermudez_PRL_2013}, but modify the way disorder acts to make it dynamic. Instead of starting with a superposition of all possible internal states and keep this quantum superposition during the experiment, one can take initially one single internal state (for example, all spins down) and then apply pseudo-random individual spin flips on different ions during the experiment. This again leads to a bimodal frequency distribution for each ion at a given time, but now the frequency becomes time dependent. In order for this to lead to transport of excitations, the differences introduced in the transverse frequencies must approximately match the gap between neighbouring ions (since the frequency distribution for each site is bimodal and not continuous).

3) An even more basic variation of 2) would be just to implement Hamiltonian (\ref{eq:frequency shift}) and initialize the spins in an alternating pattern (up, down, up, down...) for the internal state. Then one can modify the frequency differences between neighbouring ions by global spin flips.

Only the first of these three alternatives provides a continuous broadening of each resonance while it also allows one to control the correlations in the frequency variations of the different ions. Indeed, the spatial correlations are determined by the axial shape of the wavefront considered, which can be given, for instance, by a standing wave, a tightly focused beam \cite{Naegerl_1999}, or a speckle pattern \cite{Arecchi_1965, Ha_2009, Mandel_Wolf, Goodman}. Therefore we restrict to this scheme and analyze its implementation in the next Section. We note that for the case of short chains, the dynamics we study can be reproduced with simpler schemes. The proposal we develop in the following, however, has the important advantage of being scalable to longer chains.

\section{Implementation of the local noise}
\label{sec:noise}

\begin{figure}[t]
\includegraphics[width=0.85\columnwidth]{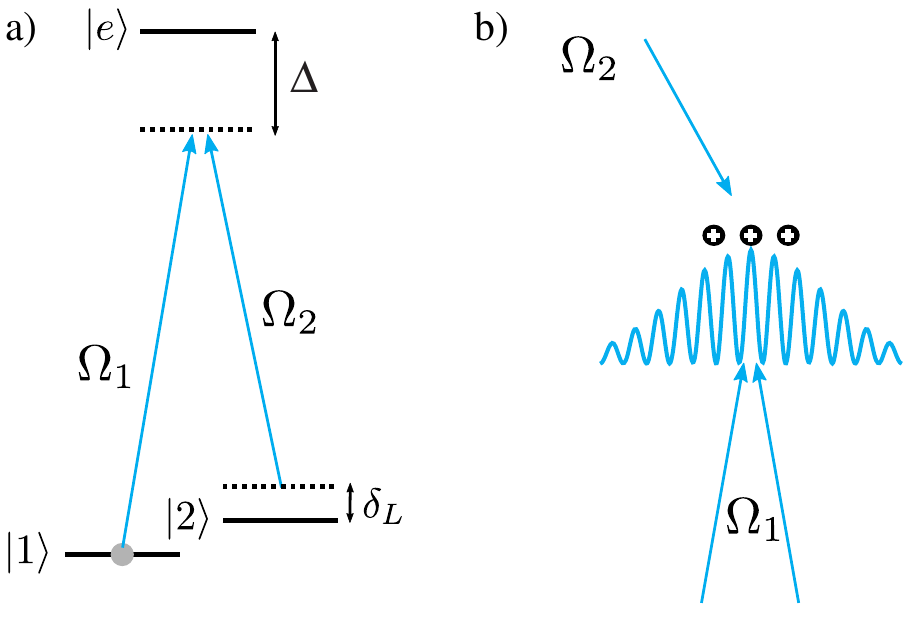}
\caption{
\label{fig:levels}
a) Scheme of the level structure required for the implementation of local transverse frequency fluctuations by means of lasers. The detunings are assumed to be large enough that all ions stay in level $\ket{1}$ at all times. b) Orientation of the lasers coupling to the two optical transitions. The transition from $\ket{1}$ to $\ket{e}$ is taken to be off-resonantly driven by a running field in the transverse direction with a wavefront of varying intensity along the axial direction. This intensity profile must be pseudo-randomly shifted, and must have spatial variations over a scale smaller than the distance between ions, while smooth enough that axial forces remain weak. The transition from $\ket{2}$ to $\ket{e}$ is driven by a single laser at an angle that can be tuned to the desired Lamb-Dicke parameter.}
\end{figure}

The scheme we consider uses a $\Lambda$ level structure with levels $\ket{1}, \ket{2}, \ket{e}$ (see Fig. \ref{fig:levels}). Here, the upper level could be a P level and the two lower levels could be S levels with a hyperfine or Zeeman splitting. Alternatively, the levels $\ket{1}$ and $\ket{2}$ could be an S and a D level respectively; our scheme only involves virtual transitions between these levels, with no need for the long coherence times required for an optical qubit. Two lasers couple each of the lower levels to 
the upper level very far off resonance so that population of the upper level remains negligible. The couplings
to the upper level have detunings of order $\Delta$, while the Rabi frequencies from the lower levels to the upper one are $\Omega_j$ (j=1,2). 
Also the indirect transition from $\ket{1}$ to
$\ket{2}$ is taken to be off-resonant so that atoms stay all the time in level $\ket{1}$, but virtual transitions affect non-negligibly the vibrational dynamics.

The field coupling the occupied level $\ket{1}$ to the upper level is assumed to correspond to a running wave propagating perpendicular to the chain, with a spatially varying wave front along the chain. The length scale in $z$ direction must be shorter than the distance between ions, but long enough that optical forces along the chain axis are weak. The laser coupling the level $\ket{2}$ to $\ket{e}$ is taken to be a running wave; the propagation direction of this laser can be chosen freely in order to obtain a convenient Lamb-Dicke parameter $\eta_x = k_x \sqrt{\hbar/(2m\omega_x)} \simeq 0.3$, large enough to have significant coupling to the transverse modes but not so large that terms of order higher than $\eta_x^2$ become relevant.

For simplicity, here and in the following we give expressions for the optical Hamiltonian acting on a single ion; the extension to several ions is straightforward. The Hamiltonian we consider is of the form:
\beq
H = \sum_{s = 1,2} \frac{\Omega_{s}}{2} [ \sigma_{+,s} e^{-i\Delta_{s}t} f_s(\vec r) + {\rm H. c.} ]
\eeq
where $s$ runs over the two lower levels, $\Delta_{s}$ is the detuning of each laser with respect to the corresponding dipolar transition, $\Omega_{s}$ gives the laser amplitude, and $\vec r$ is the position of the ion. We have assumed that each of the laser fields couples with only one of the lower levels, and we have defined the raising operators $\sigma_{+,s} = \ketbra{e}{s}$ with $s=1,2$. The Hamiltonian is written in interaction picture with respect to the electronic levels, and after a rotating wave approximation to leave out terms that rotate at twice the optical frequencies. The complex functions $f_s$ contain the spatial dependence of the fields, and read:
\begin{eqnarray}
f_1(\vec r) &=& e^{i k_{1x} x} f_z(z),\\
f_2(\vec r) &=& e^{i\vec k_2\cdot \vec r}.
\end{eqnarray}
Here, $\vec k_2$ is the wave vector of the laser coupling to level $\ket{2}$, while the field that couples to level $\ket{1}$ is the product of a running wave in the transverse direction with an axial profile given by $f_z$. We note that for arbitrary $f_z$ the function $f_1$ does not satisfy the wave equations for the optical field. We assume, however, that the axial spatial variations have a scale much larger than $\lambda_{1x}=2\pi/|k_{1x}|$, so that this factorization is a good approximation. 

Applying standard procedures to eliminate the upper level, one gets essentially the same effective Hamiltonian as in \cite{Bermudez_PRL_2013}, except for the spatial dependence of the fields. For the derivation, one needs to assume:
\beq
|\Delta| \gg |\Omega_1|, |\Omega_2|, \Gamma, |\delta_L|, |\delta_L\pm\omega_n|
\eeq
where $\Gamma$ is the decay rate from the upper level
(otherwise spontaneous decay cannot be neglected), $\delta_L$ is the detuning of the Raman transition between $\ket{1}$ and $\ket{2}$, and $\omega_n$ are the frequencies of the motional modes. 
The resulting effective 
Hamiltonian is composed of two parts, one that is diagonal with respect to the internal levels and corresponds to an ac-Stark shift,
\beq
H_1 = \proj{1} \frac{|\Omega_1|^2}{4\Delta} |f_z(z)|^2 + \proj{2} \frac{|\Omega_2|^2}{4\Delta} ,
\eeq
and one that gives a Raman coupling between internal levels:
\beq
H_2 = \frac{\Omega_1^* \Omega_2}{4\Delta} f_z(z)^* \ketbra{1}{2} e^{-i(k_x x - \delta_L t)} + {\rm H.c.} ,
\eeq
where $k_x = k_{1x}-k_{2x}$ is the transverse component of the effective wave vector of the Raman transition. In this expression, we are neglecting a dependence of the form $\exp\{ik_{2z}z\}$ assuming that it does not give rise to resonant contributions that can create axial excitations (it does cause an axial dependence of the phase of the driving, which is irrelevant for our purposes).

The coupling to the transverse position operator $x$ in $H_2$ can then be used to introduce the desired variation in the 
transverse frequencies, due to virtual processes that involve the creation and elimination of a phonon. The prefactor $f_z$ will make this frequency shift different for the different ions in the chain, and this axial profile is assumed to be shifted pseudorandomly during the experiment. Hamiltonian $H_1$ contains a term that acts like an undesired optical potential and that in general, if $\Omega_1$ and $\Omega_2$ are comparable, is much larger than the desired effect. To reduce the magnitude of this unwanted potential, we will assume that:
\beq
|\Omega_1| \ll |\Omega_2| .
\label{eq:Omegas}
\eeq
The second term in $H_1$ is a constant shift in energy that can be reabsorbed by changing to a different
rotating frame so that one gets in $H_2$ a time dependence with a detuning $\delta_L'$ instead of $\delta_L$.

We now assume that $H_2$ is sufficiently far off-resonance from all transitions, so that the dominant effect comes from 
second-order perturbation theory, which requires:
\beq
\frac{|\Omega_1 \Omega_2|}{4|\Delta|} \ll |\delta_L'|, |\delta_L'\pm \omega_n| .
\label{eq:perturbation}
\eeq
Under this assumption, considering that the ions stay at all times in level $\ket{1}$, 
and adding the remaining part of $H_1$, one obtains the following effective Hamiltonian for the motion:
\begin{multline}
H_{\rm eff} = \frac{|\Omega_1|^2}{4\Delta} |f_z(z)|^2 \Bigg\{ 1 - \frac{|\Omega_2|^2}{4\Delta} 
\Bigg[\frac{1}{\delta_L'}  + \frac{\eta_x^2}{\delta_L'+\omega_x} \\
 + \eta_x^2 a^\dagger_x a_x \left(\frac{1}{\delta_L'+\omega_x}+\frac{1}{\delta_L'-\omega_x} \right) \Bigg] \Bigg\} .
\end{multline}
This expression is achieved using an expansion in the Lamb-Dicke parameter $\eta_x$, and neglecting the coupling to axial modes since spatial variations along $z$ are assumed to be weak. 

All the terms that do not involve the mode operators can be cancelled out if the detunings and couplings satisfy the 
relation:
\beq
\label{eq:detuning condition}
4\Delta = |\Omega_2|^2 \left(\frac{1}{\delta_L'} + \frac{\eta_x^2}{\delta_L'+\omega_x} \right) .
\eeq
We note that condition (\ref{eq:Omegas}) makes it possible to satisfy this equality without breaking assumption (\ref{eq:perturbation}). When condition (\ref{eq:detuning condition}) is satisfied, we find the desired Hamiltonian:
\begin{multline}
H_{\rm eff} = - \left|\frac{\Omega_1 \Omega_2}{4\Delta}\right|^2 |f_z(z)|^2 \, \eta_x^2 \, a^\dagger_x a_x\\
 \left(\frac{1}{\delta_L'+\omega_x}+\frac{1}{\delta_L'-\omega_x} \right) 
 \label{eq:final H}
\end{multline}
which induces a shift in the transverse frequency that depends on the axial position of the ion with respect to the intensity profile that controls the intensity of the Raman transition. By pseudo-randomizing this profile, one finally obtains dephasing of the transverse modes enabling transport along a chain of previously localized modes.

Equality (\ref{eq:detuning condition}) involves a transverse frequency $\omega_x$ which will be slightly different for the different ions. 
However, the corresponding term appears multiplied by $\eta_x^2$ which is small, the differences in frequencies are also small, and  $\delta_L'$ must be larger than the trapping frequency $\omega_x$, so the effect of these frequency differences in Eq.  (\ref{eq:detuning condition}) can be neglected. It should also be noticed that the cancellation need not be perfect for the scheme to work; it is enough to require that the optical forces along the axial direction are much smaller than the Coulomb repulsion so that the positions of the ions do not vary significantly due to these forces. 

We note that the statistical properties of the frequency fluctuations experienced by the ions will strongly depend on the shape of the axial profile and the kind of variations introduced in it. We consider for instance the simple case of an axial standing wave with constant amplitude and wavelength but varying phase: in this case the fluctuations in Eq.~(\ref{eq:final H} ) will be contained between two limiting values, with a probability distribution given by the values taken by the function $\sin^2$. The fluctuations on different ions will have correlations determined by the wavelength: by choosing $\lambda_z = 8d/(2n+1)$, the fluctuations for ions at a distance $d$ will be uncorrelated, while at distance $2d$ they will be anticorrelated. Spatial correlations can be suppressed or reduced using instead a speckle pattern, which will give rise to pseudo-random spatial distributions \cite{Arecchi_1965, Ha_2009, Mandel_Wolf, Goodman}, or tightly focused beams pointing pseudo-randomly at varying ions.

A realistic implementation of our scheme could be carried out with the following parameter values:  $\omega_x\sim2\pi\times$ 400 kHz, $\Delta\sim2\pi\times$ 100~GHz,  $\Omega_2\sim2\pi\times$ 2~GHz, so that $\delta_L'\sim2\pi\times$ 10~MHz. The amplitude of the fluctuations can be tuned through $\Omega_1$; a choice of $2\pi\times$ 200~MHz gives local frequency variations between zero and about $2\pi\times$ 20 kHz. The case of a multi-isotope chain requires closer attention. We consider specifically the case of the even-numbered isotopes of Ytterbium \cite{Moehring_2007}. Consecutive isotopes have a difference in frequency for the S-P dipole transition of about 1~GHz, a difference which is negligible comparable to the detuning chosen from this transition $\Delta\sim2\pi\times$ 100~GHz. Also, by only choosing even-numbered isotopes, with no hyperfine structure, one avoids any additional complexity in the structure of the lower levels so that the presented scheme holds for all ions in the chain.

\section{Measurement}
\label{sec:measurement}

The goal of the final measurement is to observe the impact of noise on transport. To achieve this, it is enough to first introduce a driving at the frequency of an ion on one edge of the chain, let the system evolve under the dynamics including frequency fluctuations, and at the end measure the occupation of the different local modes to detect whether the motional excitation has spread over the chain.

Depending on the amount of heating present, it may happen that by the time one wants to measure the transport, heating has masked everything else. For chains with more than a few ions, initial ground-state cooling becomes also extremely demanding. To tackle these problems, it can be useful to consider a different, a bit more sophisticated approach, which allows one to ``filter out'' the purely thermal contributions to the excitations. The idea is very simple and relies on the fact that thermal noise is symmetric in phase space, i.e.~excitations introduced by thermal noise have no preferred phase. This means, if one first displaces an ion at one end of the chain and at a later time measures first moments of the form of $\langle x_j \rangle$ and $\langle p_j \rangle$, thermal noise cannot contribute to the observed value, and the signal must be entirely due to the initial excitation. 

\begin{figure*}[t]
\includegraphics[width=0.95\textwidth]{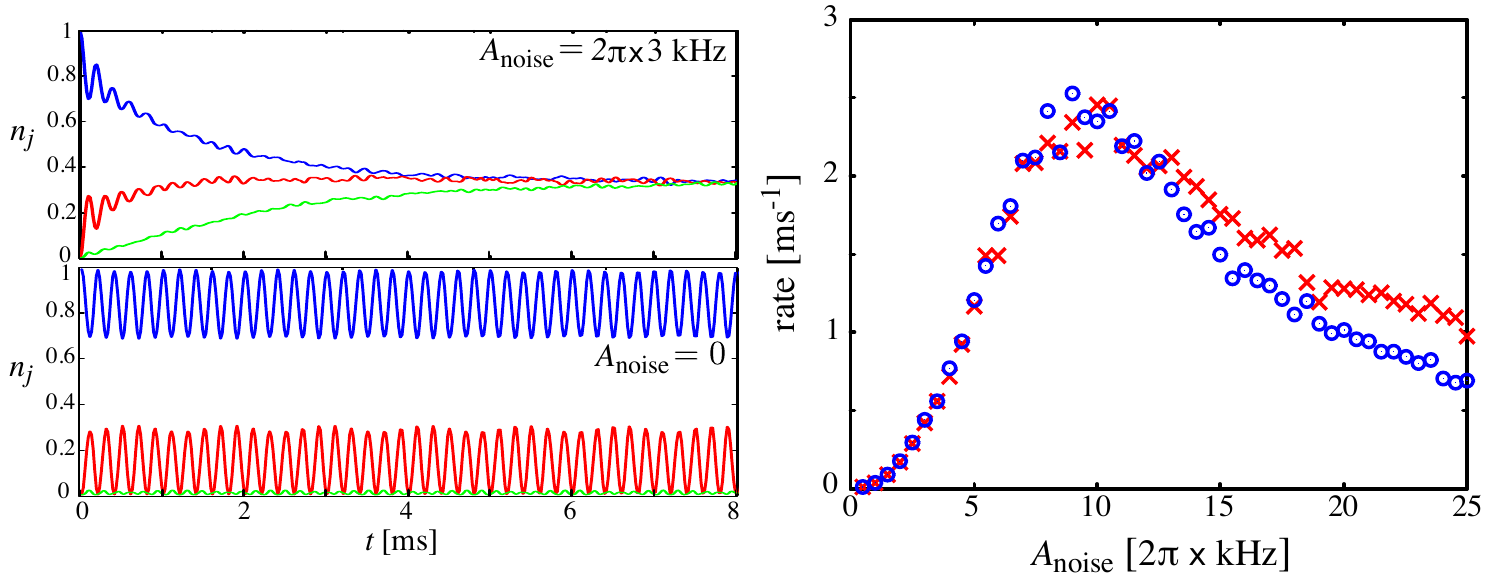}
\caption{
\label{fig:dynamics}
\textbf{Motional excitation dynamics in an inhomogeneous 3-ion chain with local dephasing.} A chain with three ions with local frequency differences of the order of  $2\pi\times$5 kHz and nearest-neighbour couplings of about $2\pi\times$1.5 kHz (more details in the text). Left panels show plots the population of sites 1 (blue), 2 (red) and 3 (green) as a function of time, when an initial excitation is placed in site 1. Bottom left: Unitary time evolution in the absence of frequency fluctuations. Top left: Frequency fluctuations are produced by a global standing wave with fluctuating phase; the noise amplitude here is taken to be $A_{\rm noise}= 2\pi\times3$ kHz (frequency shifts lie between 0 and $2 A_{\rm noise}$); the axial wavelength is taken to be $\lambda_z = 8d/(2n+1)$ with $d$ the distance between ions, so that neighbouring ions experience uncorrelated shifts. Fluctuations are left constant over a time step of $t = 20 \mu$s and then suddenly switched to another random value; the results are averaged over 600 realizations (not enough for convergence but enough to see representative behaviour). The right-hand side panel shows approximate rates for equilibration of the system for different values of the amplitude of the frequency fluctuations (blue circles). For comparison, we show with red crosses the rates found with no spatial correlations.
}
\end{figure*}

Measuring the first moments directly is not trivial in trapped ions; standard measurements of the motion involve sideband excitations and return the population of the different vibrational levels. However, it is straightforward to measure the first moments from a combination of displacement pulses and measurements of the mean occupations. Let us consider a displacement superoperator $\mathcal D (\alpha_j)$ acting on ion $j$, which acts on the annihilation operator in the form: $\mathcal D (\alpha_j) a_j = a_j + \alpha_j$. The application of such a pulse followed by a measurement of the mean excitation of site $j$ gives the result:
\begin{multline}
n_j (\alpha_j) = \langle [\mathcal D (\alpha_j) a_j]^\dagger [\mathcal D (\alpha_j) a_j] \rangle \\
= \langle a_j^\dagger a_j \rangle + |\alpha_j|^2 + 2 {\rm Re}(\alpha_j^* \langle a_j \rangle )
\end{multline}
This means that $\langle a_j \rangle$ can be extracted by combining the measurement results for different phase choices, in the form:
\begin{multline}
\langle a_j \rangle = \frac{1}{3|\alpha_j|} \Big[ n_j (|\alpha_j|) +  e^{2\pi i /3} n_j (|\alpha_j| e^{2\pi i /3}) +\\
 e^{-2\pi i /3} n_j (|\alpha_j| e^{-2\pi i /3}) \Big]
\end{multline}
where $\langle a_j \rangle$ must be interpreted as ${\rm Tr} (\rho a_j)$.

The measurement of first moments to observe the transport due to local phase fluctuations has the problem that the fluctuations themselves lead to a fast averaging out of the phase of the first moments. One can avoid this problem averaging over absolute values in the form of $\overline{|\langle a_j \rangle|^2}$, where the angle brackets indicate the quantum-mechanical mean for a given realization of the fluctuations, and the bar above stands for the statistical average over different realizations. This is equivalent to an average over mean occupation numbers excluding the contributions corresponding to thermal noise in the trap or imperfect initial ground-state cooling.

One should note that if the system is coupled to an environment which is the source of the heating, the excitations being transported will suffer from damping. However, in a typical ion setup the rate for damping of vibrations is lower by many orders of magnitude than the rate at which thermal excitations are introduced in the system, and thus the strategy proposed is an appropriate solution for the filtering of the excitations of interest.

\section{Expected behaviour for short chains}
\label{sec:results}

As a simple illustration of the ideas presented, we consider a chain of three sites. Taking parameters from \cite{Tolazzi-thesis}, we set an axial frequency $\omega_z=2\pi\times 40$ kHz, such that the nearest-neighbour couplings from Eq. (\ref{eq:couplings}) are approximately $2\pi\times$1.45 kHz. Local transverse frequencies are taken to be $2\pi\times$435, 439.5, and 445 kHz. For these parameter values the modes are strongly localized and therefore transport between sites is suppressed: Fig. \ref{fig:dynamics}, bottom-left, shows the evolution of populations of the three sites as function of time when an initial excitation is introduced in site 1; the maximum probability to find the excitation in site 3 at a later time is of about 2\%. 

In contrast, the top-left figure shows the results in the presence of frequency fluctuations due to a global standing wave as proposed in Section \ref{sec:noise}. Its amplitude is taken to be such that it induces shifts between zero and $2\pi\times6$~kHz. The fluctuations are introduced by randomly setting the phase of the standing wave, letting the system evolve by the corresponding fixed Hamiltonian for a time interval of 20 $\mu$s, and then shifting the phase again. In order to eliminate correlations between the frequency shifts experienced by neighbouring ions, we choose a ratio of $8/(2n+1)$ between the wavelength of the standing wave and the distance between ions. As can be seen in the figure, in this scenario excitations tend to become equally distributed over the three sites, with an equilibration time scale of about 5 ms. We note that the behaviour is similar to the one obtained if independent frequency fluctuations are introduced in each site.

For the Hamiltonian considered, the evolution of site populations when the initial state of site 1 is a coherent state $\ket{\alpha}$ with $|\alpha| =1$ is the same as when exactly one vibrational excitation is introduced in that site. If the initial state is a coherent state with a different value of $|\alpha|^2$, the results are simply rescaled by this value. Working with coherent states has some advantages: calculations of the evolution involving large numbers of excitations can be done efficiently using the formalism of Gaussian states \cite{Gaussian}, and coherent states can be created in the lab in a more direct way than states with a definite number of excitations. 

Although the approach to equilibrium observed in the previous figures is clearly not exponential, an exponential fit of the evolution of the population of site 3 can be used to obtain an approximate value of the equilibration rate. In the right panel of Fig.~\ref{fig:dynamics} the results of this fit are shown with blue circles, for the same parameters as the previous Figures. The rate for equilibration of the excitation among the different sites depends on the strength of the fluctuations applied, and the limits of very weak or very strong dephasing are detrimental for transport. 

The red crosses in the plot correspond to the rates obtained for a case with the same probability distribution for the frequency of each individual ion, but with no spatial correlations. The effect of spatial correlations is found to become significant for large frequency fluctuations. In this regime, the correlations of the fluctuations on sites 1 and 3 for the case of a standing wave lead to a slower equilibration compared with the uncorrelated case. This has a simple explanation: for large fluctuations, the oscillations in the evolution of populations are suppressed, and the rates for equilibration scale like the inverse of the dephasing amplitude. A standing wave with no correlations between neighbouring ions has anticorrelated fluctuations between sites 1 and 3; this can be described by an additional term in the master equation, which enhances the decay of coherences between those two sites.

\begin{figure}[ht]
\includegraphics[width=0.95\columnwidth]{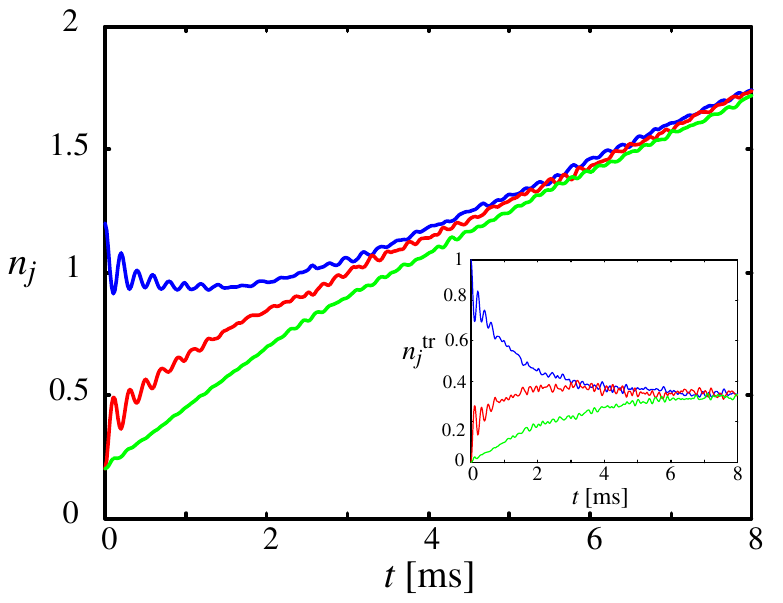}
\caption{
\label{fig:thermal}
\textbf{Effect of residual thermal excitation.} Same as in Figure \ref{fig:dynamics} (top-left) but including an initial residual occupation of 0.2 in each site and heating noise during the experiment, with an increase of 0.15 phonons per ms per ion. The inset illustrates how the proposed measurement scheme filters out the thermal behaviour; thus, the plot shows only the population of each site as consequence of the initial injected excitation being transported throughout the chain, $n_j^{\rm tr} = \overline{|\langle a_j \rangle|^2}$ (more details in the text).
}
\end{figure}

In the plots so far no thermal noise is included, and the initial state is assumed to be perfect vacuum. In practice, the motion of a trapped ion will experience thermal heating to some extent which greatly depends on the kind of trap. 
For an experiment lasting some milliseconds, one could expect a few thermal phonons \cite{Brownnutt_2015}. Also, perfect ground-state initialization can be very demanding, and a few excitations might already be present in the initial state. 
As an example, Fig. \ref{fig:thermal}
shows the same as Fig. \ref{fig:dynamics} (top-left) but including thermal noise on all sites, and an initial population of 0.2 in  each site. For simplicity, thermal noise is modeled by the standard master equation describing the coupling to a Markovian reservoir characterized by a rate 2$\kappa$ and a mean excitation number $\bar n$ corresponding to thermal equilibrium \cite{BreuerPetruccione}. We take $\bar n = 1.5 \times10^{7}$, and an increase of 0.15 phonons per ms for each ion (which gives a coupling $\kappa = 5 \times10^{-9}$~ms$^{-1}$).

In principle, the transport dynamics can be inferred from the plots in presence of thermal noise subtracting the excitations that are due to imperfect initial state preparation and heating during the experiment. Nevertheless, this requires a reliable characterization of these processes. Indeed, one should be able to prove, for instance, that the tendency observed in Fig. \ref{fig:thermal} to an equilibration of the excitations is not due to a difference in the noise rates. Alternatively, one can solve this problem performing measurements of the first moments as discussed in the previous Section. In this way one can filter out thermal excitations using the fact that thermal noise has no preferred phase. The results are shown in the inset of Fig.~\ref{fig:thermal}, where a behaviour very similar to the one of Fig.~\ref{fig:dynamics} (top-left) can be recovered.

\section{Concluding remarks}
\label{sec:conclusions}

We have shown how the use of appropriately chosen lasers can lead to local dephasing noise for the transverse vibrational degrees of freedom of ions in a chain. The method we propose is scalable, and the fluctuations introduced with this procedure take time-dependent pseudo-random values from a continuum distribution, features which represent an improvement with respect to previous proposals \cite{Bermudez_PRL_2013}. As an illustration of the idea, we consider the example of vibrations in an angle trap or a multi-isotope chain, where the variations of local trapping frequency lead to localization of the excitations. In this scenario, we show how the addition of local fluctuations can bridge the frequency gaps and allow for the tunneling of excitations. 

The results presented demonstrate how local dephasing noise suppresses localization permitting transport. This scheme can be extended to include situations with a stationary flow of excitations by combining our proposal with the methods proposed in \cite{Bermudez_PRL_2013} for the measurement of heat flows in systems which include sources and sinks. We expect that these tools will contribute to the implementation of more general simulations of quantum transport in noisy and disordered environments using ultracold trapped ions.

\section{Acknowledgements}

The authors thank F. Schmidt-Kaler and J. Ro{\ss}nagel for sharing early data and parameters from the angled Paul trap at the JGU-Mainz which inspired this work. C. T. S. acknowledges support from the Fundaci\'on Bunge y Born and from the Alexander von Humboldt Foundation.


\begin{thebibliography}{99}

\bibitem{Engel_Nature_2007}
G. S. Engel, T. R. Calhoun, E. L. Read, T.-K. Ahn, T. Mancal, Y.-C. Cheng, R. E. Blankenship, and G. R. Fleming, {Nature} {\bf 446}, 782 (2007).

\bibitem{Panitchayangkoon_PNAS_2010}
G. Panitchayangkoon, D. Hayes, K. A. Fransted, J. R. Caram, E. Harel, J. Wen, R. E. Blankenship, and G. S. Engel, {Proc. Nat. Acad. Sci. Am.} {\bf 107}, 12766 (2010).

\bibitem{Struempfer_JPCL_2012}
J. Str\"umpfer, M. \c{S}ener and K. Schulten,
{J. Phys. Chem. Lett. 3}, 536 (2012).

\bibitem{Chin_PhilTrans_2012}
A. W. Chin, S. F. Huelga and M. B. Plenio, {Phil. Trans. Royal Soc. A} {\bf 370}, 3638 (2012).

\bibitem{Chin_NatPhys_2013}
A. W. Chin, J. Prior, R. Rosenbach, F. Caycedo-Soler, S. F. Huelga and M. B. Plenio, 
{Nature Phys.} {\bf 9}, 113 (2013).

\bibitem{Mendoza-Arenas_2013}
J. J. Mendoza-Arenas, T. Grujic, D. Jaksch and S. R. Clark, 
Phys. Rev. B {\bf 87}, 235130 (2013).

\bibitem{Debora_2014}
D. Contreras-Pulido, M. Bruderer, S. F. Huelga and M. B. Plenio, 
New J. Phys. {\bf 16}, 113061
(2014).

\bibitem{Ghosh_2005}
P. K. Ghosh, D. Barik, and D. S. Ray, 
Phys. Rev. E {\bf 71}, 041107 (2005).

\bibitem{Kim_2010}
K. Kim,	M.-S. Chang,	S. Korenblit,	R. Islam,	E. E. Edwards,	J. K. Freericks,	G.-D. Lin,	L.-M. Duan, and C. Monroe, 
Nature \textbf{465}, 590 (2010).

\bibitem{Senko_2015}
C. Senko, P. Richerme, J. Smith, A. Lee, I. Cohen, A. Retzker, and C. Monroe, 
Phys. Rev. X. \textbf{5}, 021026 (2015).

\bibitem{Schindler_2013}
P. Schindler, M. M\"uller, D. Nigg, J. T. Barreiro, E. A. Martinez, M. Hennrich, T. Monz, S. Diehl, P. Zoller, and R. Blatt, 
Nature Physics \textbf{9}, 361 (2013).

\bibitem{Jurcevic_2014}
P. Jurcevic, B. P. Lanyon, P. Hauke, C. Hempel, P. Zoller, R. Blatt, and C. F. Roos
Nature \textbf{511}, 202 (2014)

\bibitem{Bohnet_2016}
J. G. Bohnet, B. C. Sawyer, J. W. Britton, M. L. Wall, A. M. Rey, M. Foss-Feig, and John J. Bollinger,  
Science \textbf{352}, 1297 (2016).

\bibitem{Clos_2015}
G. Clos, D. Porras, U. Warring, and T Schaetz, 
arXiv preprint arXiv:1509.07712 (2015).

\bibitem{Ramm_2014}
M. Ramm, T. Pruttivarasin, and H. H\"affner,
New J. Phys. \textbf{16} 063062 (2014).

\bibitem{Horak_2012}
P. Horak, A. Dantan, and M. Drewsen,
Phys. Rev. A \textbf{86}, 043435 (2012).

\bibitem{Garcia-Mata_2007}
I. Garc{\'i}a-Mata, O.V. Zhirov, and D.L. Shepelyansky, Eur. Phys. J. D {\bf 41}, 325 (2007).

\bibitem{Benassi_2011}
A. Benassi, A. Vanossi, and E. Tosatti,
Nat. Commun. {\bf 2}, 236 (2011).

\bibitem{Bylinskii_2015}
A. Bylinskii, D. Gangloff, V. Vuleti\'{c},
Science \textbf{348}, 1115 (2015).

\bibitem{Pruttivarasin_2011}
T. Pruttivarasin, M. Ramm, I. Talukdar, A. Kreuter and H. H\"affner, 
New Journal of Physics \textbf{13}, 075012 (2011).

\bibitem{Baltrusch_2011}
J. D. Baltrusch, C. Cormick, G. De Chiara, T. Calarco, and G. Morigi,
Phys. Rev. A \textbf{84}, 063821 (2011).

\bibitem{Fogarty_2016}
T. Fogarty, H. Landa, C. Cormick, and G. Morigi,
Phys. Rev. A \textbf{94}, 023844 (2016)

\bibitem{Schlawin_2014}
F. Schlawin, M. Gessner, S. Mukamel, and A. Buchleitner, 
Phys. Rev. A \textbf{90}, 023603 (2014).

\bibitem{Gessner_2014}
M. Gessner, F. Schlawin, H. H\"affner, S. Mukamel, and A. Buchleitner,  
New Journal of Physics \textbf{16}, 092001 (2014).

\bibitem{Lemmer_2015}
A. Lemmer, C. Cormick, C. T. Schmiegelow, F. Schmidt-Kaler, and M. B. Plenio, 
Phys. Rev. Lett. \textbf{114}, 073001 (2015).

\bibitem{Schmiegelow_2016}
C.T. Schmiegelow, H. Kaufmann, T. Ruster, J. Schulz, V. Kaushal, M. Hettrich, F. Schmidt-Kaler, and U.G. Poschinger, 
Phys. Rev. Lett. \textbf{116}, 033002 (2016).

\bibitem{Freitas_2015}
N. Freitas, E. A. Martinez, and J. P. Paz, 
Phys. Scr. {\bf 91}, 013007 (2015).

\bibitem{Bermudez_PRL_2013}
A. Bermudez, M. Bruderer and M.~B. Plenio, Phys. Rev. Lett. {\bf 111}, 040601 (2013).

\bibitem{heat-engine-1}
O. Abah, J. Ro{\ss}nagel, G. Jacob, S. Deffner, F. Schmidt-Kaler, K. Singer, and E. Lutz, 
Phys. Rev. Lett {\bf 109}, 203006 (2012).

\bibitem{heat-engine-2} 
J. Ro{\ss}nagel, O. Abah, F. Schmidt-Kaler, K. Singer, and E. Lutz, 
 Phys. Rev. Lett. {\bf 112}, 030602 (2014).
 
\bibitem{heat-engine-3}  
 J. Ro{\ss}nagel, S. Dawkins, N. Tolazzi, O. Abah, E. Lutz, F. Schmidt-Kaler, and K. Singer,
Science \textbf{352}, 325 (2016).

\bibitem{Leibfried_2003}
D. Leibfried, R. Blatt, C. Monroe, and D. Wineland,
Rev. Mod. Phys. {\bf 75}, 281 (2003).

\bibitem{Tolazzi-thesis} K. N. Tolazzi, Diploma Thesis, Johannes Gutenberg Universit\"at Mainz (2014).

\bibitem{Morigi_Walther_2001}
G. Morigi and H. Walther,
E.P.J. D {\bf 13}, 261 (2001).

\bibitem{Brownnutt_2015}
M. Brownnutt, M. Kumph, P. Rabl, and R. Blatt,
Rev. Mod. Phys. \textbf{87}, 1419 (2015).

\bibitem{Naegerl_1999}
H. C. N\"agerl, D. Leibfried, H. Rohde, G. Thalhammer, J. Eschner, F. Schmidt-Kaler, and R. Blatt,
Phys. Rev. A {\bf 60}, 145 (1999).

\bibitem{Arecchi_1965}
F. T. Arecchi, 
Phys. Rev. Lett. {\bf 15}, 912 (1965).

\bibitem{Ha_2009}
W. Ha, S. Lee, Y. Jung, J.~K. Kim, and K. Oh, 
Opt. Exp. {\bf 17}, 17536 (2009).

\bibitem{Mandel_Wolf}
L. Mandel and E. Wolf, Optical coherence and quantum optics (Cambridge University Press, 1995).

\bibitem{Goodman}
J. W. Goodman, Speckle phenomena in optics: theory and applications (Roberts and Company, 2007).

\bibitem{Moehring_2007}
D. Moehring, ``Remote Entanglement of Trapped Atomic Ions''. Ph.D. Thesis, Univ. of Maryland (2007).

\bibitem{Gaussian}
C. Weedbrook, S. Pirandola, R. Garc\'ia-Patr\'on, N. J. Cerf, T. C. Ralph, J. H. Shapiro, and S. Lloyd,
Rev. Mod. Phys. {\bf 84}, 621 (2012).

\bibitem{BreuerPetruccione}
H.-P. Breuer and F. Petruccione,
The Theory of Open Quantum Systems.
(Oxford University Press, 2007).

\end{thebibliography}
\end{document}